\documentclass{iopconfser}
\usepackage{graphicx}
\usepackage{adjustbox}
\usepackage{float}
\usepackage{array}
\usepackage{longtable}
\usepackage{subcaption}
\usepackage{xcolor}
\usepackage{url}

\begin{document}

\title{Comparative Analysis of Machine Learning based Intrusion Detection in Realistic IoT Networks}

\author{Rana Alharbi, Chuadhry Mujeeb Ahmed, Newcastle University, UK}
\email{R.A.A.Alharbi2@newcastle.ac.uk}

\vspace{1.5cm}

\begin{abstract}
The Internet of Things (IoT) is rapidly growing and expanding into various sectors, such as healthcare, transportation, smart homes, and more. Despite the benefits of using IoT devices, they present several challenges. Given the significant role these devices play in our lives, it is crucial to address issues related to their security and privacy. These devices are limited in resources, which complicates their security and the protection of the data that they manage. The paper aims to examine intrusion detection systems using the Gotham2025 dataset, generated through the Gotham testbed, which consists of 78 emulated IoT devices utilising various protocols, including MQTT, CoAP, and RTSP, to assist in safeguarding IoT networks from attacks. We conduct a comparative analysis between five machine learning algorithms, including Random Forest, XGBoost, Logistic Regression, Naive Bayes, and Deep Neural Network. We demonstrate that the Random Forest Classifier was the top-performing model, achieving an F1-score of 0.99 in classifying attacks.
\end{abstract}

\section{Introduction}
The term Internet of Things is defined as a network of devices, enabling information exchange online \cite{Alqarawi03072023}. The idea of connecting devices other than standard computers to the Internet has existed for a long time. For instance, an internet connected toaster was introduced in 1990. The term Internet of Things was originally introduced in 1998 by Kevin Ashton \cite{Bandyopadhyay2011}. Furthermore, the number of internet connected devices is growing rapidly. Estimates by Cisco suggest that by 2030, 500 billion devices will be connected to the Internet \cite{s21041174}. The Internet of Things is rapidly gaining popularity and expanding into various sectors, including healthcare, transportation, smart homes, and more. It includes the use of baby and home monitoring cameras, video-enabled doorbells, autonomous vehicles, smartwatches, and so on \cite{10.1007/s11277-021-09116-5}. However, this advancement presents challenges. Despite the benefits associated with increased usage, risks concerning security and privacy arise, which make safeguarding these IoT networks a vital priority. Part of these safeguarding techniques is the use of machine learning. 

Machine Learning (ML) techniques are gaining popularity in security \cite{7123563,cpss_Ahmed2020}. It primarily relies on automating activities based on inferred knowledge from the data it utilises. From attack detection to authentication and access control in IoT networks, malware analysis, anomaly and intrusion detection, and IoT device identification, machine learning is creating a more secure IoT environment \cite{9060970,scan_cycle}. Previous studies have considered old datasets and tested with a single ML technique. A comprehensive comparison between different ML techniques and the latest device dataset is missing.  Our paper aims to compare a broad range of machine learning based intrusion detection systems using the latest Gotham2025 dataset \cite{belarbi2025gothamdataset2025reproducible}, generated through the Gotham testbed\cite{10049670}, which consists of 78 emulated IoT devices utilising various protocols, including MQTT, CoAP, and RTSP, to help safeguard IoT networks from attacks. The remainder of the document is organised as follows: 
\begin{itemize}
    \item Section 2 (Background and Related Work): Discusses the background and related work, including vulnerabilities in the Internet of Things, datasets, and intrusion detection systems. 
    \item Section 3 (Methodology) describes the methodology, characteristics of the dataset, experimental setup, selected machine learning models, and evaluation metrics.
    \item Section 4 (Experimental Results) presents the experiments and discusses the results of each evaluated machine learning model and comparatively assesses model performance.
    \item Section 5 (Conclusions and Future Work) concludes the paper and outlines directions for future work.
\end{itemize}

\section{Background and Related Work}

This section highlights IoT vulnerabilities, explains datasets and why they are needed to detect these threats and introduces Intrusion Detection Systems.

\subsection{Internet of Things Vulnerabilities}
Inadequate security of IoT devices leads to disastrous security incidents that can affect organisations' reputation and finances. Furthermore, they may result in the disclosure of users’ sensitive data or data destruction \cite{HUMAYUN2021105}. A significant attack took place in 2016 involving the Mirai botnet, which targeted IoT devices and utilised them to attack the Domain Name Server Dyn in a Distributed Denial of Service (DDoS). Antonakakis et al. conducted an analysis of the Mirai botnet attack and concluded that the inadequate security of IoT devices is the primary cause of the attack \cite{203628,limitations_kalman_filter}.

These vulnerabilities include unauthorised access, open ports, and unprotected network interfaces \cite{KAUR2023100780}. Outdated firmware is another vulnerability in IoT networks, where many IoT devices fail to update to the latest versions, rendering them susceptible to attacks due to this insecure firmware. Additionally, the use of weak passwords, particularly in situations where an organisation uses the same weak passwords across multiple devices, can lead to one compromised device, causing a chain of compromised devices. Attackers often exploit the limited resources of IoT devices, making them one of their primary targets \cite{SHAHIN2024102685}. Due to constraints in storage and power capabilities, these devices typically require lightweight encryption methods to mitigate vulnerabilities such as exposed communications and flaws in the data transmission processes of IoT devices \cite{Elrawy2018,SHAHIN2024102685}.

Attacks exploiting these vulnerabilities typically appear as denial-of-service (DoS) attacks, where devices are overwhelmed with requests, affecting their availability. It can be accomplished with the help of network protocols such as UDP or ICMP flood attacks, ping of death, and SYN/ACK attacks. On the other hand, Distributed Denial-of-Service (DDoS) has the same impact, utilising multiple entities to attack the same target. Brute force is another attack that attempts unauthorised access by guessing weak passwords \cite{SHAHIN2024102685}. Network scanning, which includes the detection of open ports or exposed network interfaces, represents another threat to IoT networks \cite{KAUR2023100780}.

What makes securing the Internet of Things a challenging process is the nature of the devices that constitute an Internet of Things network. These networks may include a variety of devices, ranging from small temperature sensors that measure room temperature and communicate with other devices to cameras or self-driving vehicles \cite{https://doi.org/10.1002/dac.5613}. These IoT networks differ in size, the number of connected devices, their energy consumption levels, the operating system used, and the protocols employed~\cite{10.1007/s11277-021-09116-5}. This heterogeneity, arising from IoT devices differing in size, capabilities, characteristics, hardware constraints, and communication protocols \cite{7123563}, presents one of the key challenges in maintaining IoT security \cite{Bandyopadhyay2011,Gautham_Challenges}.

\subsection{Datasets}
The increasing adoption of IoT devices across various sectors has led researchers to create datasets that can be utilised to train machine learning models for effectively detecting different intrusions and attacks against these networks, thus enhancing performance when these models are applied in real-world scenarios. The effectiveness of the model is heavily influenced by the nature and quality of the dataset. Datasets are created using testbeds that simulate IoT environments, and they can be generated by creating realistic IoT environments with actual IoT devices. Different data collection methods have multiple factors, including network size, the number of devices utilised, the type of conducted attacks, and the total duration of device traffic collection. The types of devices included in the testbed impact the dataset's quality, especially as technology evolves and real-world devices change. Another critical factor is that the broader the functionalities offered by an IoT device, the wider the attack vector on that IoT network \cite{KAUR2023100780}. 

One of the most popular datasets for testing Intrusion Detection Systems (IDS) in cybersecurity is KDD-99, created in 1990 and improved on by NL-KDD in 2009. Most research has relied on this dataset to evaluate their IDS. However, this dataset does not represent emerging threats and vulnerabilities, which means that testing on an outdated dataset will not accurately reflect how these IDS perform when deployed in the real world. Consequently, the IDS will not generalise well to new threats and attacks \cite{s23052415}. Other datasets tailored for IoT Intrusion Detection Systems include: MedBIoT, IoTID20, IoT-23 and CICIoT2023 \cite{UA._OYELAKIN_2023}. 

For a dataset to be effective, it must include a variety of new attacks and threats utilising real IoT devices to create a realistic network environment, as some datasets employ traditional computers instead of IoT devices to simulate attacks \cite{s23135941}. Among the recent datasets are CICIoT2023\cite{s23135941} and Gotham2025 \cite{belarbi2025gothamdataset2025reproducible} Datasets. The Gotham2025 Dataset is designed to assist research in large-scale IoT networks. It provides heterogeneity by offering various types of devices, protocols, and attack surfaces to simulate real-world IoT environments. The nature of the dataset mimics real-world scenarios by collecting data in separate files rather than dividing data after collection, which some researchers do to test federated learning approaches \cite{belarbi2025gothamdataset2025reproducible}. 

\subsection{Intrusion Detection Systems (IDS)}
Traditional IDS, such as signature-based systems, effectively detects known attacks. However, they struggle to identify new or unseen threats. Machine learning and deep learning models have become the foundation of many modern IDSs, successfully addressing the limitations of traditional systems. Key challenges in implementing IDS include associated computational costs, particularly as we deal with IoT devices that have limited capabilities. We must reduce these costs while ensuring high accuracy and robustness. Another challenge lies in the nature of the dataset, which often presents a skewed distribution, and the choice of relevant features in intrusion detection, which in turn influences model complexity~\cite{umer2021attack}. Furthermore, the appropriate selection of features significantly impacts training time and IDS performance \cite{RAHMAN2025100082}. Machine learning techniques are valuable for identifying patterns in data and detecting suspicious behaviour, whether using supervised learning models with labelled data or unsupervised ones, which lack labels and is useful for identifying zero-day attacks. However, it can produce many false positives. Deep learning and reinforcement learning are among the methods employed in IDS implementation \cite{s23052415}. 

\section{Methodology}
\begin{figure}
    \centering
    \includegraphics[width=0.4\linewidth]{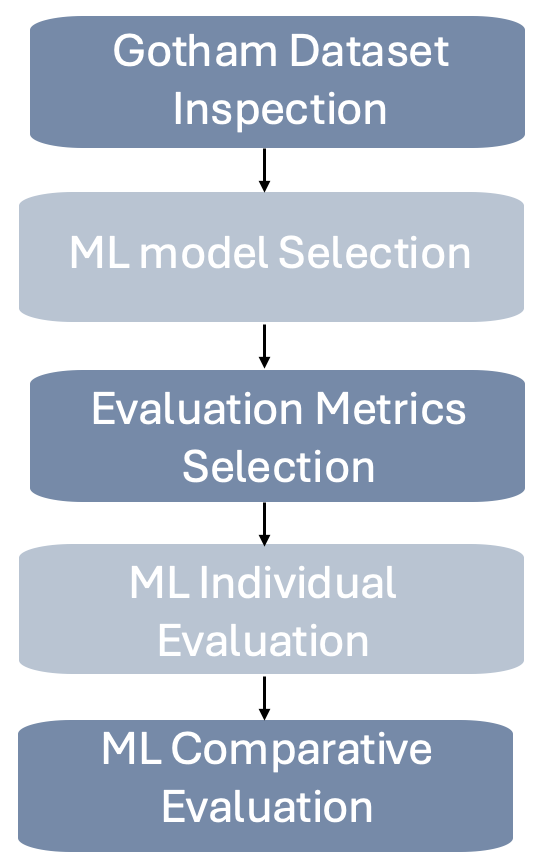}
    \caption{Overview of the Methodology.}
    \label{fig:Methodology}
\end{figure}

To achieve the objectives of this study, we follow the methodology in Figure~\ref{fig:Methodology}: we first inspect the Gotham dataset, select the machine learning algorithms, choose the evaluation metrics, evaluate each machine learning model individually and comparatively analyze the performance of the selected ML models. This chapter outlines the methodology and gives a detailed explanation of the Gotham dataset,  the preprocessing steps,  the model selection process, the evaluation metrics used, and the experimental setup designed to assess model performance.
\subsection{Dataset and Preprocessing}
\label{sec:dataset}
We used the GothamDataset2025 \cite{belarbi2025gothamdataset2025reproducible}, which contains labelled network traffic data representing a diverse range of attack scenarios and normal traffic. The normal traffic is collected over two hours, whereas each attack is gathered over a period of one to one and a half hours. They collected traffic separately, ensuring that the data for each device was reflected in its own file to facilitate decentralized analysis and learning. Since the data is spread over multiple files and we are working in a centralised setting, we combined all traffic into a single dataframe to evaluate Machine Learning performance.

    \begin{table}[ht]
    \centering
     \small
    \begin{tabular}{|c|c|}
        \hline
        \textbf{Label} & \textbf{Count} \\
        \hline
        Benign & 12256883 \\
        \hline
        Mirai UDP Flooding & 8897895 \\
        \hline
        Mirai TCP Flooding & 6548173 \\
        \hline
        Mirai GRE Flooding & 5911401 \\
        \hline
        TCP Scan & 737764 \\
        \hline
        CoAP Amplification & 274837 \\
        \hline
        Telnet Brute Force & 227649 \\
        \hline
        Merlin TCP Flooding & 120000 \\
        \hline
        Merlin ICMP Flooding & 57580 \\
        \hline
        Merlin UDP Flooding & 29996 \\
        \hline
        Merlin C\&C Communication & 29356 \\
        \hline
        Ingress Tool Transfer & 21587 \\
        \hline
        Unknown & 7670 \\
        \hline
        File Download & 7196 \\
        \hline
        UDP Scan & 4242 \\
        \hline
        Mirai C\&C Communication & 1074 \\
        \hline
        C\&C Communication & 528 \\
        \hline
        Reporting & 450 \\
        \hline
        \textbf{Total} & 35134281 \\
        \hline
    \end{tabular}
    \caption{Dataset before grouping.}
    \label{tab:Before_G}
    \end{table}

   \begin{table}[ht]
    \centering
     \small
    \begin{tabular}{|c|c|}
        \hline
        \textbf{Label} & \textbf{Count} \\
        \hline
         Brute Force & 227649  \\
        \hline
         C\&C Communication & 30958 \\
        \hline
         DoS & 21839882 \\
        \hline
         Infection & 29233 \\
        \hline
         Network Scanning & 742006 \\         
        \hline
         Normal & 12256883  \\
        \hline
        \textbf{Total} & 35161611 \\
        \hline
    \end{tabular}
    \caption{Dataset after grouping.}
    \label{tab:After_G}
    \end{table}

\begin{figure}
    \centering
    \includegraphics[width=1\linewidth]{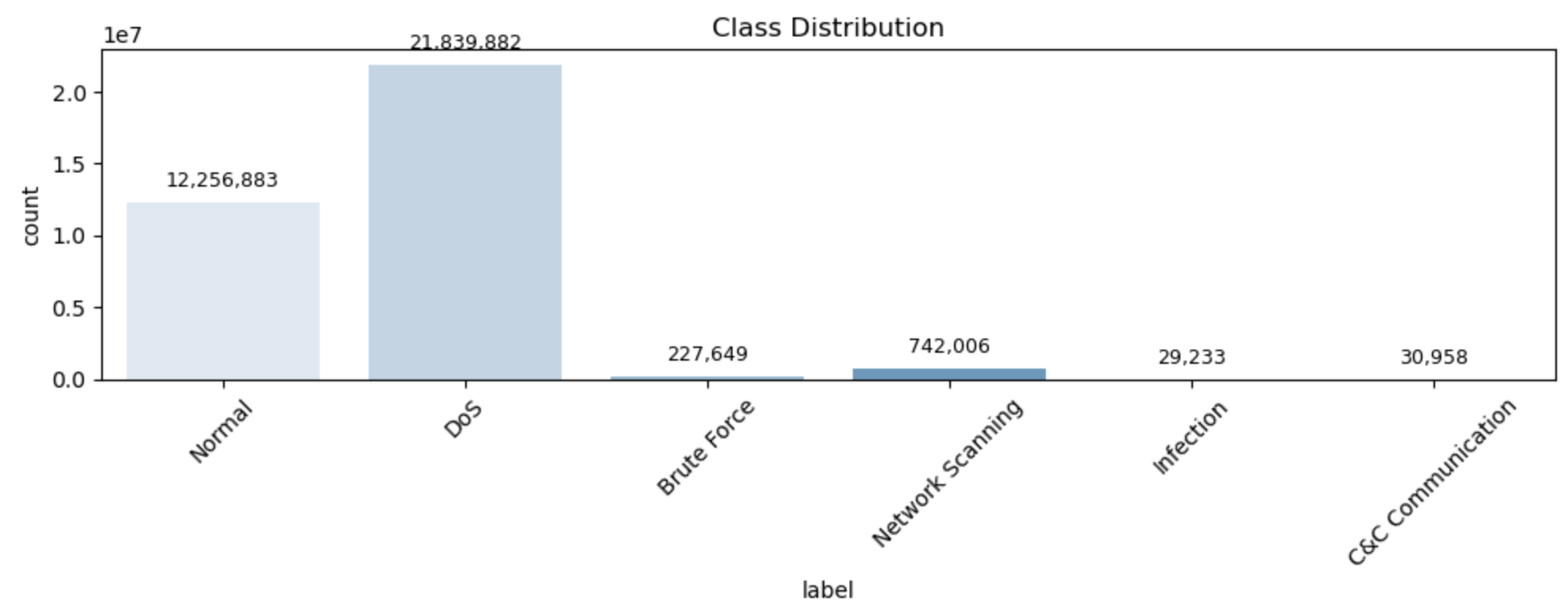}
    \caption{Class Distribution for Dataset.}
    \label{fig:ClassDist}
\end{figure}

The type of attacks collected in the Gotham Dataset 2025 are grouped into: \textbf{Normal, Network Scanning, Brute Force, Infection, C\&C Communication, DoS.}
Table~\ref{tab:Before_G} shows attacks before grouping. The dataset is unbalanced. Moreover, grouping the attacks into six classes made DOS attacks dominant in the number of items across the training data. Figure~\ref{fig:ClassDist} shows the class breakdown after the data have been grouped into six labels, and Table~\ref{tab:After_G} shows the number of attacks after the grouping. The preprocessing was performed using the author's code for the dataset provided on GitHub \cite{belarbi2025gothamdataset2025reproducible}. Their code serves as a preprocessing step that includes data splitting, feature extraction, along encoding to ensure consistent data handling across both their deep learning methods and our machine learning approaches. These preprocessing steps applied to the Gotham dataset include extracting protocol types and source and destination ports from the frame. Protocols feature, converting ports into categorical values, separating IP flags and TCP flags into individual binary features, standardizing the timestamp, handling missing values, scaling numerical features, and detailed attack labels are mapped into six categorical labels, which are then encoded into numerical values suitable for training machine learning models. Some adjustments were made to the code, as some traffic was incorrectly categorised during preprocessing; for instance, benign traffic and certain C\&C communications were labelled as 'Other' or unknown traffic.

The Gotham dataset is tailored to smart city environments and was collected using the Gotham testbed \cite{10049670}. It includes emulated traffic from various device types, such as air quality and building monitors, which include readings from fifteen and twenty-seven sensors, respectively, including humidity, temperature, and other types of sensors. Other device types include cooler motors, which use acceleration and speed sensors to monitor fan vibration. These devices communicate using several lightweight IoT protocols: Message Queuing Telemetry Transport (MQTT) \cite{mqtt5}, a standard lightweight messaging protocol for devices with limited resources, Constrained Applications Protocol (CoAP protocol) \cite{rfc7252} is used for machine-to-machine communications such as building automation and smart energy, and Real Time Streaming Protocol (RTSP) \cite{rfc2326} which is used in streaming audio and video. The Gotham dataset is relatively new and has not been explored. It introduces heterogeneity in the types of devices included in the smart city environment and generates revenue to facilitate Federated learning approaches that limit data sharing and protect privacy due to the manner in which this traffic is collected.

\subsection{Selected Machine Learning Algorithms}
\label{sec:MLSel}
The selection of the algorithm was based on its relevance and impact in the field. Random Forest was chosen due to its widespread use and recognition as one of the most cited algorithms in the literature. Although XGBoost and Logistic Regression have fewer citations, they still demonstrate relevance in Intrusion Detection Systems \cite{RAHMAN2025100082}. Naive Bayes Classifier was also included, as it has been utilised in various IoT security solutions \cite{KAUR2023100780}. 
\begin{itemize}
\item \textbf{Random Forest Classifiers:} Random forest is one of the ensemble methods in which the final decision of the classifier is based on the predictions of multiple machine learning algorithms. The Random Forest consists of a collection of decision trees, where the dataset is divided into various subsets. Each decision tree takes a section of the features and some records in a process called Bootstrap. Each tree makes a decision based on the subset of the assigned dataset. The final decision of the Random Forest is determined by a majority vote, where the most frequently chosen classification becomes the final prediction in a process known as aggregation. Bootstrap and aggregation together are referred to as bagging. An alternative method of training on different subsets is known as pasting. Using multiple decision trees helps to mitigate the problem of high variance present in a single decision tree \cite{10.5555/3378999}.

\item \textbf{XGBoost Classifier:}
XGBoost is a machine learning technique that stands for Extreme Gradient Boosting, which is an optimised version of Gradient Boosting. It offers speed, scalability, and portability. It involves training predictors in a sequential manner, where each predictor adjusts its predecessor's error and fits it to the residual errors of the previous predictor. XGBoost prevents overfitting by incorporating early stopping in its techniques \cite{10.5555/3378999}.

\item \textbf{Logistic Regression:}
Logistic Regression is based on calculating the probability that an instance belongs to a particular class. It handles multiclass classification by producing a vector of probabilities. The softmax function is applied, which gives a probability distribution over classes. For instance, given an input x, it computes a score for each class, where the class with the highest probability represents the predicted class \cite{10.5555/3378999}.
 
\item \textbf{Naive Bayes Classifier:}

Naive Bayes is similar to Bayesian classifiers, but it simplifies the complexity present in them. To reduce this complexity, Naive Bayes employs a conditional independence assumption, which assumes that there is no dependence among all features describing x given the class label. This assumption decreases the number of parameters, thus lowering the complexity \cite{mitchell2017chapter3}. It relies on Bayes' theorem, which updates the probability distribution after observing some data \cite{10.5555/3378999}. 

\end{itemize}

\subsection{Evaluation Metrics}
\label{sec:metrics}
Evaluation metrics are designed for Machine Learning and Deep Learning algorithms performance assessment. The expected behaviour of models is to achieve high True Positives (when a model makes a positive prediction and it is accurate) and high True Negatives (when a model makes a negative prediction and it is accurate) while minimising False Positives (when a model makes a positive prediction that is incorrect) and False Negatives (when a model makes a negative prediction that is incorrect). The application of each metric depends on the specific use case in which it is employed. Accuracy alone is inadequate for evaluating models, particularly when the datasets are skewed. For instance, in a binary classification scenario, if class A has 90\% of the data samples while class B accounts for only 10\%, a classifier that predicts all instances as class A would achieve an accuracy of 90\%. However, this percentage does not provide a good indication of how the classifier behaves when it is employed in the real world. For instance, when we aim to reduce false positives, we should prioritise precision. On the other hand, when we want to decrease false negatives, we should concentrate on recall. Various metrics can provide differing insights into the model performance \cite{10.5555/3378999}. The models discussed in section ~\ref{sec:MLSel} will be evaluated using the different metric evaluations, including Accuracy, precision, and recall, with the focus on F1-Score since we are working with an unbalanced dataset. Confusion matrices are utilised as well to show false positives and false negatives that exist in each class. The metrics employed in this paper\cite{s23135941}:

\noindent \textbf{Accuracy} metric refers to the percentage of correct predictions from total predictions, both correct and incorrect, see Equation~\ref{eq:accuracy}.

\begin{equation}
Accuracy = \frac{TP + TN}{TP + TN + FP + FN}
\label{eq:accuracy}
\end{equation}

\noindent\textbf{Precision}  refers to positive prediction accuracy, which indicates how many of the predicted positive instances were correct, as shown in Equation~\ref{eq:pre}.
\begin{equation}
Precision = \frac{TP}{TP+FP}
\label{eq:pre}
\end{equation}

\noindent\textbf{Recall}, (True Positive Rate and sensitivity), indicates how many of the positive instances the model correctly identifies, as shown in Equation~\ref{eq:recall}.
\begin{equation}
\label{eq:recall}
Recall = \frac{TP}{TP+FN}
\end{equation}

\noindent\textbf{F1-Score}, consider both false negative and false positive using precision and recall as shown in Equation~\ref{eq:f1}. 

\begin{equation}
F1 = 2 \times \frac{Precision \times Recall}{Precision+Recall}
\label{eq:f1}
\end{equation}


\noindent The \textbf{confusion matrix} is a good indicator of how many instances are being misclassified. The matrix shows a count of each class and how many of these instances are being misclassified. The columns in the matrix represent the predicted class, where the rows represent the actual class. 

\subsection{Experimental Setup}
All of the machine learning models were implemented using the Scikit-learn \cite{pedregosa2011scikit}. The Machine Learning models utilised in this paper include Random Forest Classifiers, XGBoost Classifier, Logistic Regression and Naive Bayes Classifier \cite{RAHMAN2025100082} \cite{KAUR2023100780}. SVM was also among the machine learning models chosen, as it ranks second to Random Forest \cite{RAHMAN2025100082}. However, it struggled to handle the large dataset effectively, as shown by our experiments, where training took days without completion on both our local computing systems and Google Colab. Training on our local system equipped with a 2.3 GHz 8-Core Intel Core i9 processor and 16 GB RAM took considerable time to run various experiments. After conducting experiments locally, we purchased a paid subscription to Google Colab Pro, which improved performance and reduced the time needed to run different experiments. Several challenges were encountered concerning interrupted experiments and device overheating during training, which impacted the experimental process.

\section{Results}
In this section, we will discuss the performance of each selected model in classifying the various attack classes using the evaluation metrics discussed in section~\ref{sec:metrics}: accuracy, precision, recall, and the resulting confusion matrix. Following the individual analysis, we conduct a comparative evaluation to identify the top-performing model.

\subsection{Random Forest Classifier}
\begin{figure}[H] 
    \centering
    \includegraphics[width=0.75\linewidth]{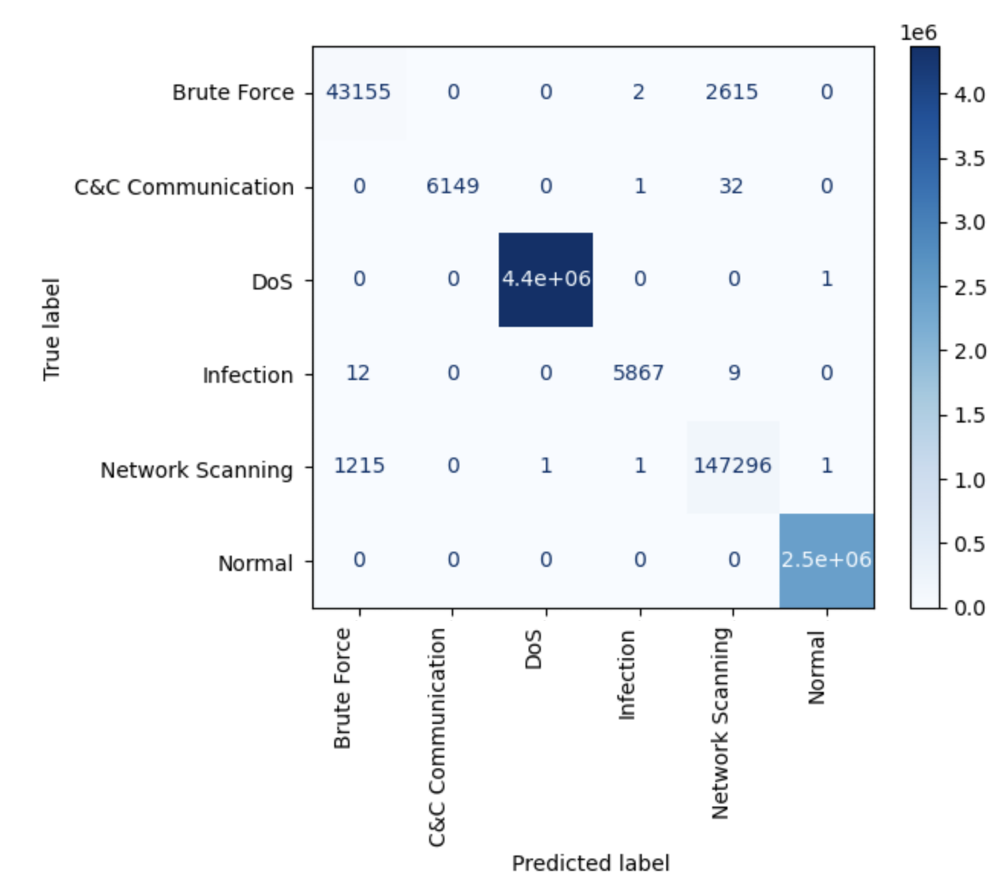}
    \caption{Random Forest Classifier Confusion Matrix}
    \label{fig:RFconf_matrix}
\end{figure}

The Random Forest Classifier performs well with the dataset, achieving nearly perfect predictions for four of the six classes. As illustrated in the confusion matrix in Figure~\ref{fig:RFconf_matrix}, Normal, C\&C Communications, Infection, and DoS demonstrate near-perfect predictions, with at most 0.45\%, which is a small percentage misclassified as other labels. However, Network Scanning and Brute Force show some confusion. The classification report in Table~\ref{tab:rf_classification} indicates an overall accuracy of 100\% and a macro-average F1-Score of 0.99, suggesting that Random Forest is effective with unbalanced data and robust in classifying network activities.

\begin{table}[H]
\centering
\small
\begin{tabular}{lcccc}
\hline
\textbf{Class} & \textbf{Precision} & \textbf{Recall} & \textbf{F1-Score} & \textbf{Support} \\
\hline
 0 Brute Force         & 0.97 & 0.94 & 0.96 & 45,772 \\
 1 C\&C Communication  & 1.00 & 0.99 & 1.00 & 6,182 \\
 2 DoS                 & 1.00 & 1.00 & 1.00 & 4,367,650 \\
 3 Infection           & 1.00 & 1.00 & 1.00 & 5,888 \\
 4 Network Scanning    & 0.98 & 0.99 & 0.99 & 148,514 \\
 5 Normal              & 1.00 & 1.00 & 1.00 & 2,451,315 \\
\hline
\textbf{Accuracy}         &       &       & \textbf{1.00} & 7,025,321 \\
\textbf{Macro Avg}        & 0.99 & 0.99 & 0.99 & 7,025,321 \\
\textbf{Weighted Avg}     & 1.00 & 1.00 & 1.00 & 7,025,321 \\
\hline
\end{tabular}
\caption{Classification Report (Random Forest Classifier)}
\label{tab:rf_classification}
\end{table}

\subsection{XGBoost Classifier}
\begin{figure}[H]
    \centering
    \includegraphics[width=0.75\linewidth]{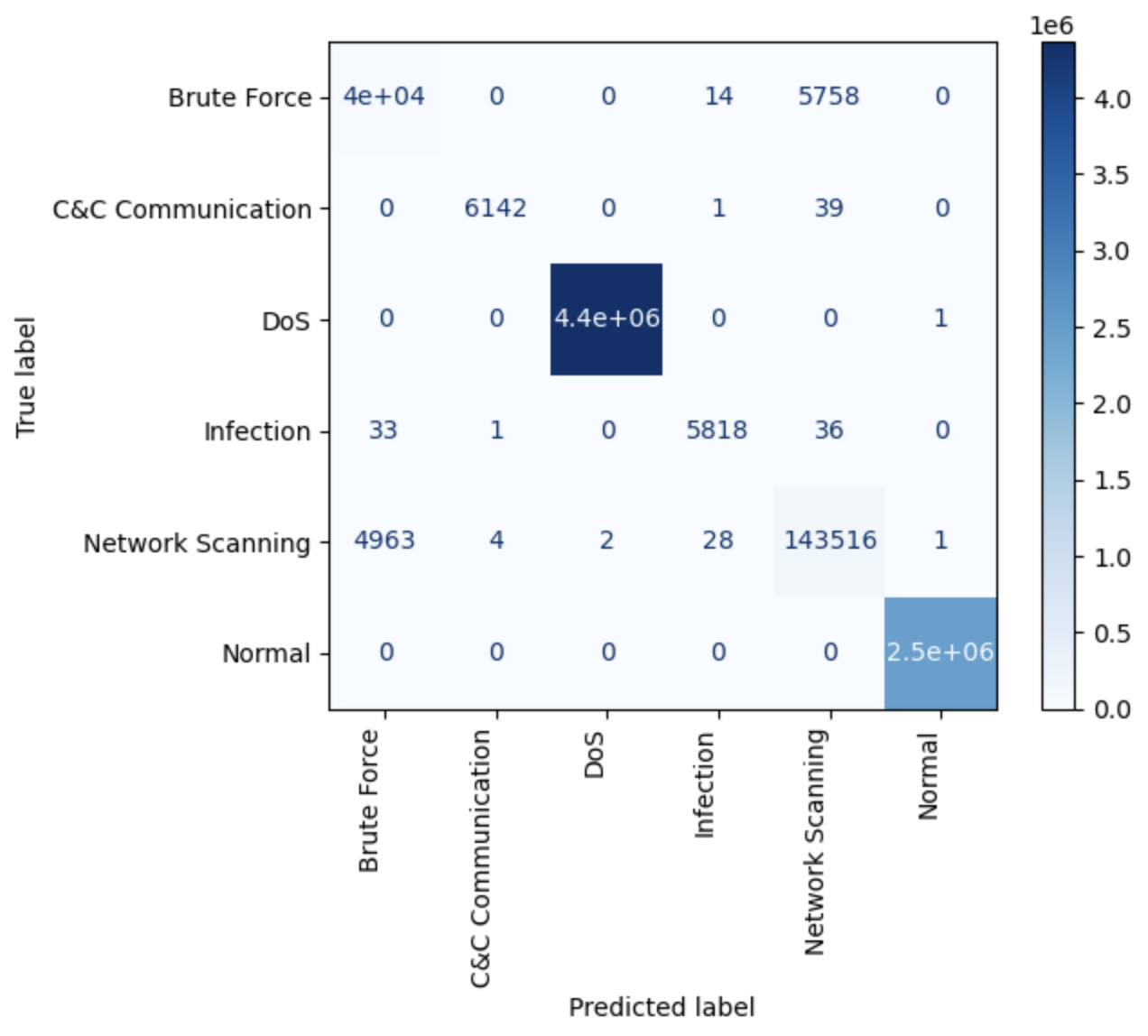}
    \caption{XGB Classifier Confusion Matrix}
    \label{fig:xgbconf_matrix}
\end{figure}

The XGBoost Classifier performs slightly worse than the Random Forest Classifier. As illustrated in the confusion matrix in Figure~\ref{fig:xgbconf_matrix}, normal traffic, C\&Cs communications, and DoS demonstrate some confusion, with at most five instances misclassified as other labels. However, network scanning is the least accurate among all six classes regarding misclassification. The classification report in Table~\ref{tab:XGB_classification} indicates an overall accuracy of 100\% and a macro-average F1-Score of 0.97, suggesting that the XGBClassifier is less effective than Random Forest with unbalanced data and in the robustness of classifying network activities.
\begin{table}[H]
\centering

\begin{tabular}{lcccc}
\hline
\textbf{Class} & \textbf{Precision} & \textbf{Recall} & \textbf{F1-Score} & \textbf{Support} \\
\hline
 0 Brute Force           & 0.89 & 0.87 & 0.88 & 45,772 \\
 1 C\&C Communication    & 1.00 & 0.99 & 1.00 & 6,182 \\
 2 DoS                   & 1.00 & 1.00 & 1.00 & 4,367,650 \\
 3 Infection             & 0.99 & 0.99 & 0.99 & 5,888 \\
 4 Network Scanning      & 0.96 & 0.97 & 0.96 & 148,514 \\
 5 Normal                & 1.00 & 1.00 & 1.00 & 2,451,315 \\
\hline
\textbf{Accuracy}       &       &       & \textbf{1.00} & 7,025,321 \\
\textbf{Macro Avg}      & 0.97  & 0.97  & 0.97 & 7,025,321 \\
\textbf{Weighted Avg}   & 1.00  & 1.00  & 1.00 & 7,025,321 \\
\hline
\end{tabular}
\caption{Classification Report for XGBClassifier}
\label{tab:XGB_classification}
\end{table}

\subsection{Logistic Regression}
\begin{figure}[H]
    \centering
    \includegraphics[width=0.75\linewidth]{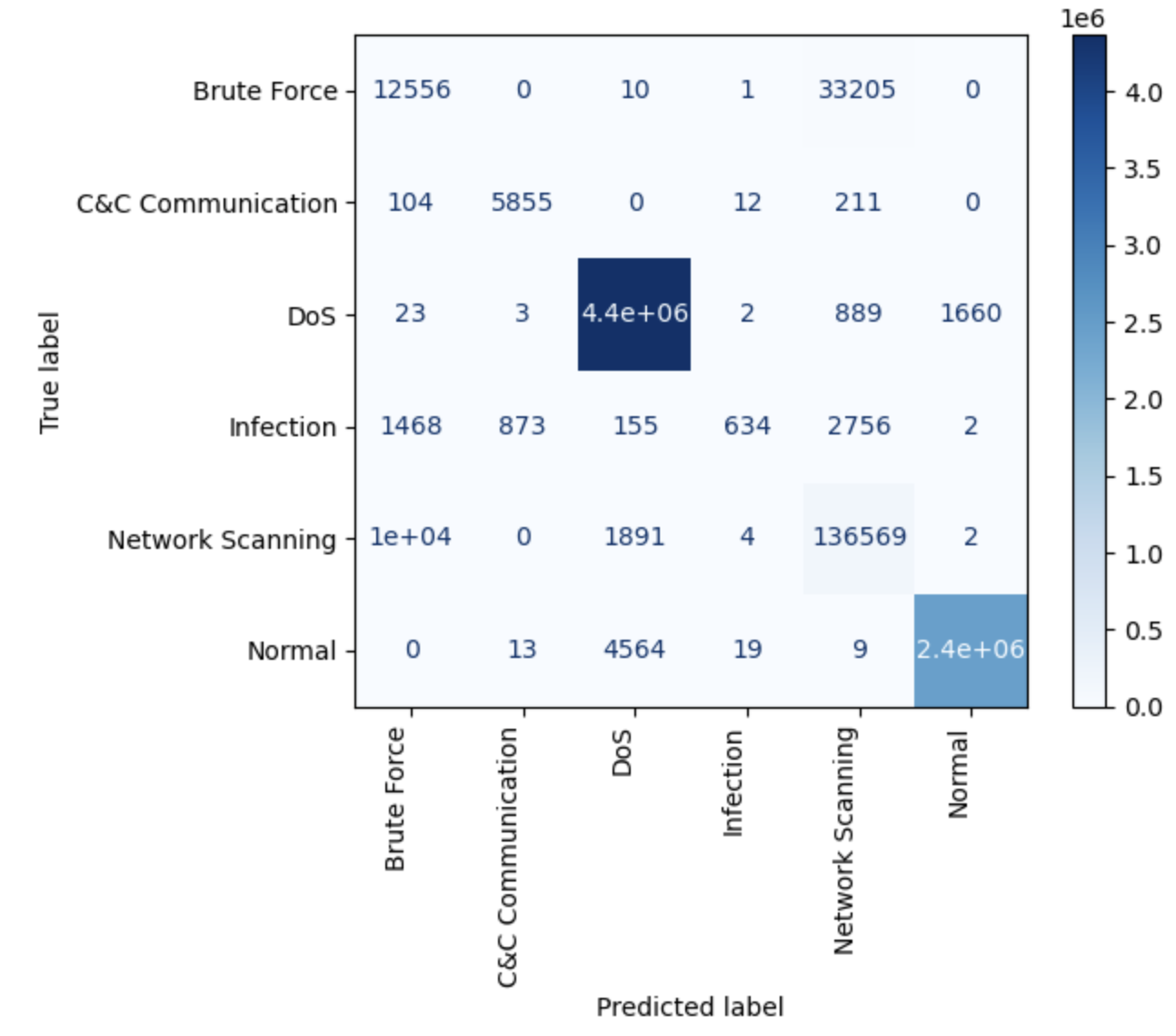}
    \caption{Logistic Regression Confusion Matrix}
    \label{fig:LRconf_matrix}
\end{figure}

The Logistic Regression model ranked second to last among the tested machine learning models. As illustrated in the confusion matrix in Figure~\ref{fig:LRconf_matrix}, Brute Force, Network Scanning, Infection, and C\&C Communication are frequently confused with other labels and are misclassified. However, Normal traffic and DoS perform better, indicating that logistic regression does not handle class imbalance well. The classification report in Table~\ref{tab:lr_classification} shows an overall accuracy of 99\% and a macro-average F1-score of 0.72. However, accuracy is not particularly relevant in this context. As presented in the table, Brute Force has an F1-score of 0.36, while Infection has an F1-score of 0.19, and Normal traffic and DoS achieve an F1-score of 1. This suggests that Logistic Regression performs well with classes that have a larger number of instances and poorly with minority classes. Nonetheless, the confusion matrix in figure~\ref{fig:RFconf_matrix} demonstrates that despite DoS achieving an F1-score of 1, the model still misclassifies 2574 instances as Brute Force, Infection, Network Scanning, and Normal traffic. This indicates that even the class with many instances is still subject to misclassification.

\begin{table}[h!]
\centering
\begin{tabular}{lcccc}
\hline
\textbf{Class} & \textbf{Precision} & \textbf{Recall} & \textbf{F1-Score} & \textbf{Support} \\
\hline
 0 Brute Force          & 0.52 & 0.27 & 0.36 & 45,772 \\
 1 C\&C Communication   & 0.87 & 0.95 & 0.91 & 6,182 \\
 2 DoS                  & 1.00 & 1.00 & 1.00 & 4,367,650 \\
 3 Infection            & 0.94 & 0.11 & 0.19 & 5,888 \\
 4 Network Scanning     & 0.79 & 0.92 & 0.85 & 148,514 \\
 5 Normal               & 1.00 & 1.00 & 1.00 & 2,451,315 \\
\hline
\textbf{Accuracy}       &       &       & \textbf{0.99} & 7,025,321 \\
\textbf{Macro Avg}      & 0.85  & 0.71  & 0.72 & 7,025,321 \\
\textbf{Weighted Avg}   & 0.99  & 0.99  & 0.99 & 7,025,321 \\
\hline
\end{tabular}
\caption{Classification Report for (Logistic Regression)}
\label{tab:lr_classification}
\end{table}

\subsection{Naive Bayes Classifier}

\begin{figure}[H]
    \centering
    \includegraphics[width=0.75\linewidth]{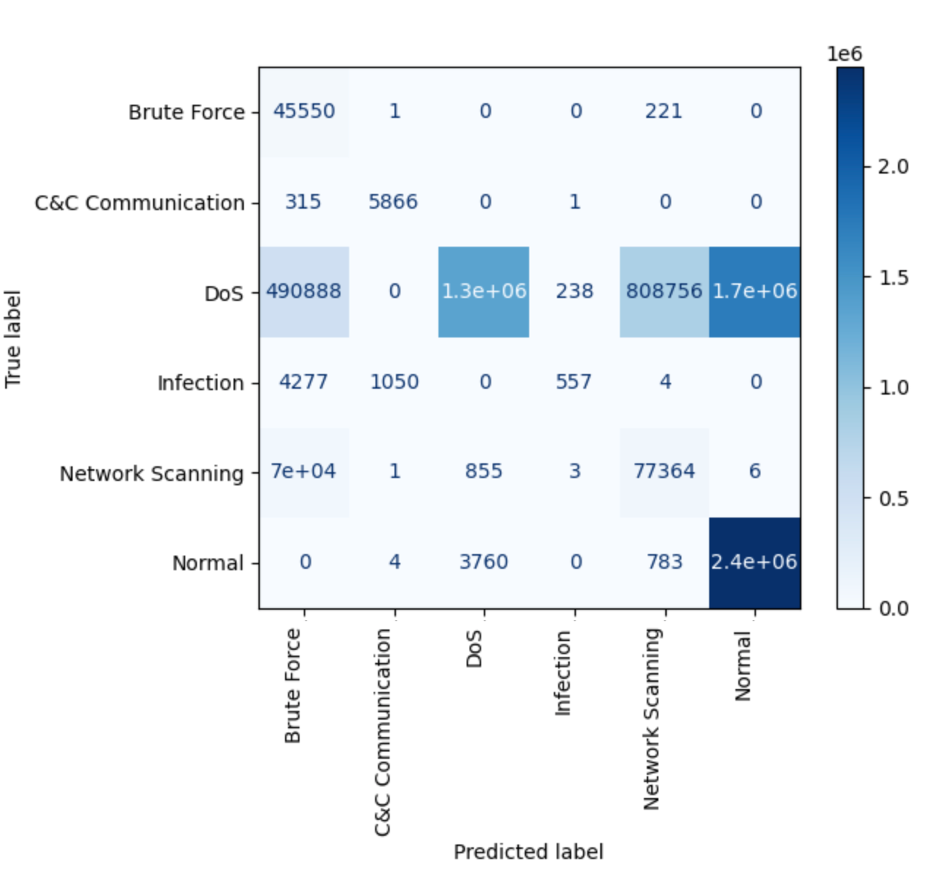}
    \caption{Naive Bayes Confusion Matrix}
    \label{fig:NBconf_matrix}
\end{figure}

Among the tested machine learning algorithms, the Naive Bayes Classifier demonstrated the lowest overall performance. The model showed inconsistent behaviour across all classes, as illustrated in the confusion matrix in Figure~\ref{fig:NBconf_matrix}. Dos is often misclassified as other classes of attacks but is frequently mistaken for normal traffic. The classification report in Table~\ref{tab:NB_classification} shows no clear correlation between the number of samples and the precision, recall and F1-score of each class. For instance, C\&C Communications, with 6,182 samples, achieved the highest F1-score, while Normal traffic and DoS, with 2,451,315 and 4,367,650 samples, respectively, had lower F1-score. The report indicates an overall accuracy of 56\%, a macro-average F1-Score of 0.43 and a weighted average F1-score of 0.55. These results suggest that the Naive Bayes classifier is ineffective for the Gotham dataset in distinguishing these attacks.

\begin{table}[h!]
\centering
\small
\begin{tabular}{lcccc}
\hline
\textbf{Class } & \textbf{Precision} & \textbf{Recall} & \textbf{F1-Score} & \textbf{Support} \\
\hline
0 Brute Force & 0.07 & 1.00 & 0.14 & 45,772 \\
1 C\&C Communication & 0.85 & 0.95 & 0.90 & 6,182 \\
2 DoS    & 1.00 & 0.31 & 0.47 & 4,367,650 \\
3 Infection & 0.70 & 0.09 & 0.17 & 5,888 \\
4 Network Scanning & 0.09 & 0.52 & 0.15 & 148,514 \\
5 Normal & 0.59 & 1.00 & 0.74 & 2,451,315 \\
\hline
\textbf{Accuracy} &  &  & 0.56 & 7,025,321 \\
\textbf{Macro Avg} & 0.55 & 0.64 & 0.43 & 7,025,321 \\
\textbf{Weighted Avg} & 0.83 & 0.56 & 0.55 & 7,025,321 \\
\hline
\end{tabular}
\caption{Classification report (Naive Bayes Classifier)}
\label{tab:NB_classification}
\end{table}

\subsection{Centralised Deep Neural Network}

\begin{figure}[H]
    \centering
    \includegraphics[width=0.75\linewidth]{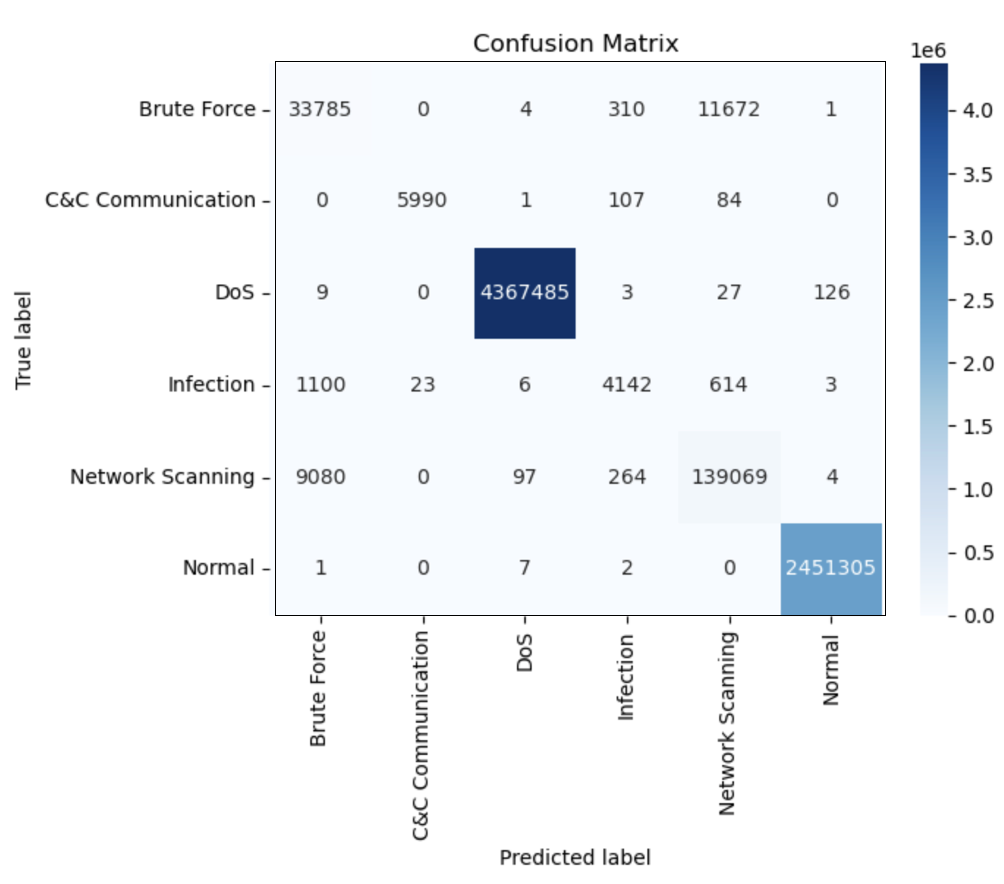}
    \caption{Deep Neural Network (DNN) Confusion Matrix}
    \label{fig:DNNconf_matrix}
\end{figure}

Researchers in \cite{belarbi2025gothamdataset2025reproducible} implemented a simple fully connected deep neural network with three hidden layers and a ReLU activation function. Table~\ref{tab:dnn_classification} indicates that the deep neural network model achieved 0.91\% as a Macro Average F1-Score, with the DoS and Normal traffic classes achieving an F1-Score of 1. The lowest scores among all six classes were observed in the Brute Force and Infection classes. The confusion matrix in Figure~\ref{fig:DNNconf_matrix} illustrates that many instances are being misclassified, with the class that has the fewest instances misclassified being Normal traffic, followed by DoS, which correlates with the large number of instances.

\begin{table}[h!]
\centering
\small
\begin{tabular}{lcccc}
\hline
\textbf{Class} & \textbf{Precision} & \textbf{Recall} & \textbf{F1-score} & \textbf{Support} \\
\hline
0 Brute Force           & 0.77 & 0.74 & 0.75 & 45,772 \\
1 C\&C Communication    & 1.00 & 0.97 & 0.98 & 6,182 \\
2 DoS                   & 1.00 & 1.00 & 1.00 & 4,367,650 \\
3 Infection             & 0.86 & 0.70 & 0.77 & 5,888 \\
4 Network Scanning      & 0.92 & 0.94 & 0.93 & 148,514 \\
5 Normal                & 1.00 & 1.00 & 1.00 & 2,451,315 \\
\hline
\textbf{Accuracy} &       &       & 1.00 & 7,025,321 \\
\textbf{Macro avg} & 0.92 & 0.89 & 0.91 & 7,025,321 \\
\textbf{Weighted avg} & 1.00 & 1.00 & 1.00 & 7,025,321 \\
\hline
\end{tabular}
\caption{Classification report (Deep Neural Network)}
\label{tab:dnn_classification}
\end{table}

\newpage
\subsection{Comparative Evaluation of Model Performance}
\begin{figure}[H]
    \centering
    \includegraphics[width=0.6\linewidth]{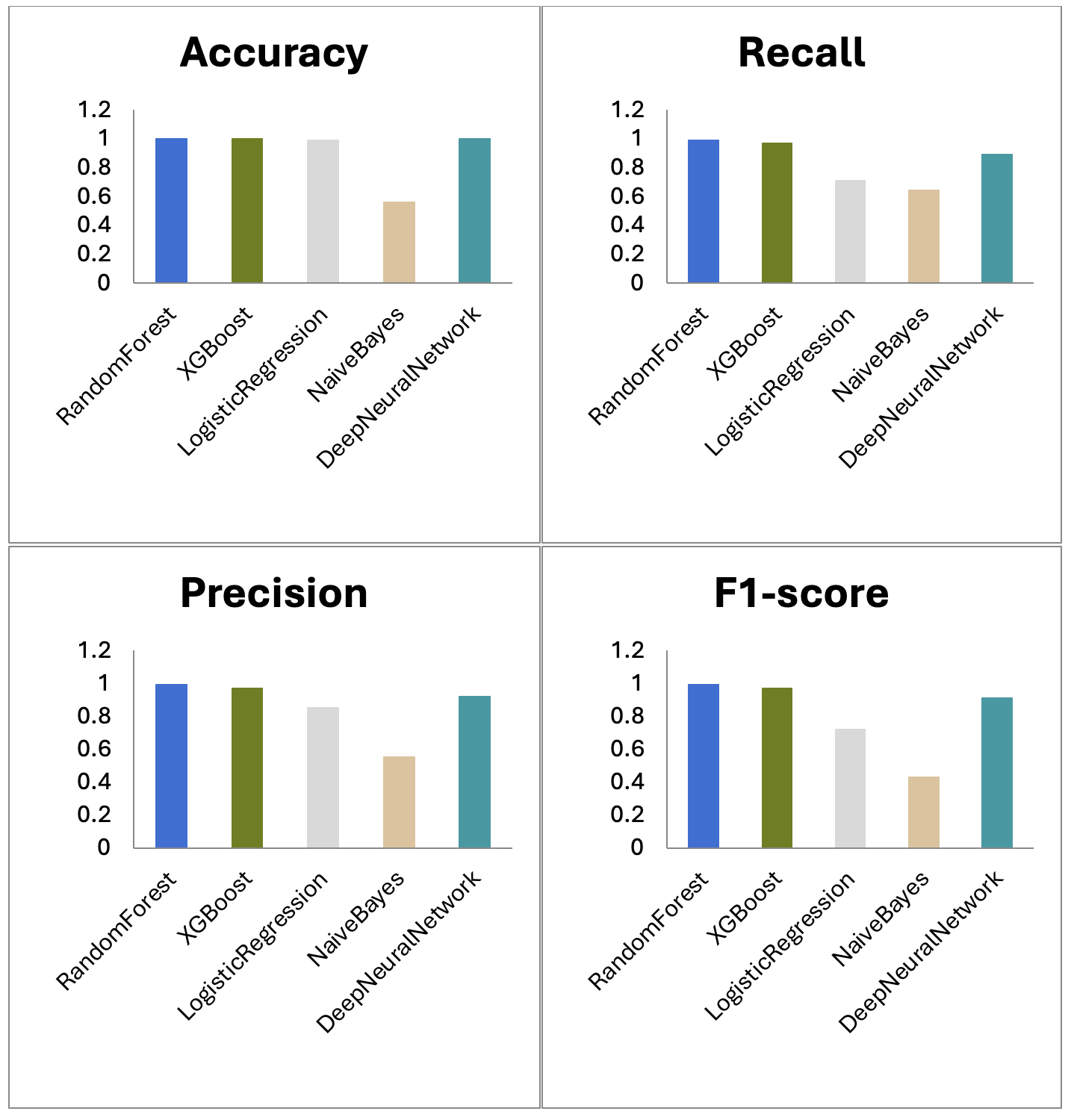}
    \caption{Models Performance.}
    \label{fig:ModelsPerformance}
\end{figure}

The performance across the five different algorithms is summarised in Table~\ref{tab:f1-accuracy-all-models} and Figure~\ref{fig:ModelsPerformance}, revealing variations in performance across classes. Random Forest and XGBoost were among the top-performing models. The random forest model achieved the highest overall accuracy and F1 score, with near-perfect classification performance for almost four out of the six classes, demonstrating its effectiveness in the Gotham dataset. XGBoost followed, achieving perfect F1 scores in three out of the six classes. A deep neural network performs well, earning an overall F1-score of 91\%. However, it underperforms in two attack classes. In contrast, Logistic Regression and Naive Bayes had the lowest performance, indicating limitations in attack classification. Logistic Regression performed better than Naive Bayes, achieving an overall F1-score of 72\% and an accuracy of 99\%. Naive Bayes, On the other hand,  achieved an accuracy of 56\%, a macro F1-score of 43\%, and a weighted F1-score of 55\%. It records the lowest performance per class, with 14\%, 15\%, and 17\% F1-scores demonstrating how it is ineffective in distinguishing these attacks.

\begin{table}[h!]
\centering
\begin{tabular}{lcccccc}
\hline
\textbf{Class} & \textbf{RF} & \textbf{XGBoost} & \textbf{DNN} & \textbf{LR} & \textbf{NB} \\
\hline
0 Brute Force        & 0.96 & 0.88 & 0.75 & 0.36 & 0.14 \\
1 C\&C Communication & 1.00 & 1.00 & 0.98 & 0.91 & 0.90 \\
2 DoS                & 1.00 & 1.00 & 1.00 & 1.00 & 0.47 \\
3 Infection          & 1.00 & 0.99 & 0.77 & 0.19 & 0.17 \\
4 Network Scanning   & 0.99 & 0.96 & 0.93 & 0.85 & 0.15 \\
5 Normal             & 1.00 & 1.00 & 1.00 & 1.00 & 0.74 \\
\hline
\textbf{Accuracy}    & 1.00 & 1.00 & 1.00 & 0.99 & 0.56 \\
\textbf{Macro F1}    & 0.99 & 0.97 & 0.91 & 0.72 & 0.43 \\
\textbf{Weighted F1} & 1.00 & 1.00 & 1.00 & 0.99 & 0.55 \\
\hline
\end{tabular}
\caption{F1-scores across different tested models.}
\label{tab:f1-accuracy-all-models}
\end{table}
\newpage
\section{Discussion}

This study utilizes the Gotham dataset, which is based on a smart city environment that has not been previously studied. It focuses on network heterogeneity and large-scale IoT networks. It has 22 features; the model's limited feature set helps lower computational demands. The dataset relies on distributed data collection, where each device maintains its own file to enable federated learning. However, the study has certain limitations. First, this work depends on supervised learning, which may struggle to identify new attack types not included in the training dataset. Such models could have difficulty generalising to unseen data. Another potential limitation is the fact that the Gotham dataset is highly imbalanced, which can cause models to be biased and struggle to identify minority classes when deployed. Finally, these models may not operate effectively on small IoT devices due to their high computational cost. Therefore, this limits the applicability of these models in resource-constrained environments.

IoT devices are typically limited in resources, restricting their ability to run complex models; these models must be efficient in terms of memory, storage, and power consumption. For example, DNNs have a large number of parameters that need to be stored for effective decision-making, along with the memory required for operational processes \cite{10.1145/3570955}. Additionally, there is a trade-off between accuracy and execution time, especially for IoT problems; for instance, achieving high accuracy might come at the cost of longer execution times. Therefore, models must make quick predictions and respond quickly, particularly in time-sensitive IoT systems \cite{Vakili2020PerformanceAA}. For example, SVM was one of the explored models, where it exhausted resources during training, where no model was produced; for that exact reason, Vakili et al. \cite{Vakili2020PerformanceAA} excluded SVM from their experiments due to its high execution time. Theoretically, SVM has a worst-case time complexity of $O(n^3)$ where the execution time scales as the cube of the number of data points $n$, which justifies the long training periods required for such models. On the other hand, Random Forest and XGBoost have a time complexity of $O(k \log n)$, where the execution time is proportional to the logarithm of the input size, making them significantly faster than SVM. Logistic regression and Naive Bayes come in the middle with linear time complexity $O(nd)$, which means their runtime grows linearly with input size \cite{han2025airegulatoryaffairsbalancing}. Han et al. \cite{han2025airegulatoryaffairsbalancing} observed that inference times of these models when deployed in the real world match their theoretical Big-O complexity. They concluded that Logistic regression is efficient in resource-constrained environments, while Naive Bayes has low inference time. However, our study indicates that fast execution times are less relevant when these two models had the lowest accuracy, precision, recall, and F1-score compared to the others. This emphasises the importance of balancing accuracy, lightweight design, and fast, reliable responses suitable for IoT devices.

\section{Conclusions and Future Work}

We present a comparative analysis of machine learning based intrusion detection systems using the Gotham2025 dataset \cite{belarbi2025gothamdataset2025reproducible} to help safeguard IoT networks from attacks. The goal is to understand the performance of standard algorithms on realistic and recent IoT data. We evaluated ML algorithms in terms of Accuracy, Precision, Recall, and F1-score as evaluation metrics using the Gotham dataset individually and conducted a comparative analysis between the five machine learning algorithms: Random Forest, XGBoost, Logistic Regression, Naïve Bayes, and Deep Neural Network. Our results show that Random Forest outperforms the other models, achieving a high Macro-Average F1 Score, which demonstrates its effectiveness in detecting attacks on IoT-specific data.

All the settings of our experiments in which these machine learning models are tested are centralised. Recently, the field of Federated Learning has mainly been utilised to preserve privacy \cite{electronics14061176}. Instead of collecting data from all devices in a centralised data storage, it allows distributed training among IoT devices, which protects privacy by eliminating the need to share their data \cite{belarbi2025gothamdataset2025reproducible}. For Future work, we intend to examine the potential of testing machine learning models and exploring other deep learning models in federated settings. 

\newpage
\bibliographystyle{unsrt}
\bibliography{references}

\end{document}